# The Biological Metaphor of a Second-Order Observer and the Sociological Discourse

Loet Leydesdorff
University of Amsterdam, Amsterdam School of Communications Research (ASCoR)
Kloveniersburgwal 48, 1012 CX  Amsterdam, The Netherlands
loet@leydesdorff.net ; http://www.leydesdorff.net

**Structured abstract; conceptual paper**

**Purpose**
In the tradition of Spencer Brown's (1969) *Laws of Form*, observation was defined in Luhmann's (1984) social systems theory as the designation of a distinction. In the sociological design, however, the designation specifies only a category for the observation. The distinction between observation and expectation enables the sociologist to appreciate the processing of meaning in social systems.

**Design**
The specification of "the observer" in the tradition of systems theory is analyzed in historical detail. Inconsistencies and differences in perspectives are explicated, and the specificity of human language is further specified. The processing of meaning in social systems adds another layer to the communication.

**Findings**
Reflexivity about the different perspectives of participant observers and an external observer is fundamental to the sociological discourse. The ranges of possible observations from different perspectives can be considered as second-order observations or equivalently as the specification of an uncertainty in the observations. This specification of an uncertainty provides us with an expectation. The expectation can be provided with (one or more) values by observations. The significance of observations can be tested when the expectations are properly specified.

**Value**
The expectations (second-order observations) are structured and therefore systemic attributes to the discourse. However, the metaphor of a (meta-)biological observer has disturbed the translation of social systems theory into sociological discourse. Different discourses specify other expectations about possible observations. By specifying second-order observations as expectations, social systems theory and sociocybernetics can combine the constructivist with an empirical approach.

**Keywords**: sociocybernetics, meaning, observer, social systems, observational report, expectation, communication

## 1. Introduction

Sociocybernetics and systems theory have often been dismissed by sociologists with the argument that the prevailing metaphor in these traditions has remained meta-biological (Giddens, 1984; Grathoff, 1978; Habermas, 1987; Habermas & Luhmann, 1971). In an important collection of articles on the subject entitled *Sociocybernetic Paradoxes* and edited by Felix Geyer and Hans van der Zouwen, Luhmann (1986, at p. 172) addressed this problem when he formulated that "because it is tied to life as a mode of self-reproduction of autopoietic systems, the theory of autopoiesis does not really attain the level of general systems theory which includes brains and machines, psychic systems and social systems, societies and short interactions." He proposed considering the processing of meaning as another, that is, non-biological, form of autopoietic organization, which he formulated as follows:

> The concept of autopoietic closure itself requires this theoretical decision, and leads to a sharp distinction between *meaning* and *life* as different forms of autopoietic organization; and meaning-using systems again have to be distinguished according to whether they use *consciousness* or *communication* as modes of meaning-based reproduction. (…) The general theory (of autopoiesis), however, is meaningful only if this implementation succeeds, because otherwise we would be unable to determine which kinds of attributes are really general. (Luhmann, 1986: 173).

It will be argued in this contribution that Luhmann was successful in achieving this "implementation." He elaborated how meaning-processing systems are different from living systems. However, in using "observation" without sufficient reflection on the status of this concept in sociological discourse, Luhmann's theory of social systems remained vulnerable to the epistemological critique of using a meta-biological metaphor (Habermas, 1987; Leydesdorff, 2000).

The clarification of this issue is urgent because the confusion has been used as an argument against sociocybernetic approaches and in favour of action theory and its derivatives in sociology (e.g., Habermas, 1981; Giddens, 1984; Münch, 1982/1988; Beck *et al.*, 2003). While observation as an operation can be considered as action, the reflexive entertaining of expectations is based on perceptions and experience. The introduction of a "second-order observer" at the level of social systems has not solved the issue, but led to philosophical discussions that distract from an empirical orientation in research (e.g., Luhmann *et al.*, 1990; Baecker, 1999; Fuchs, 2004). The systems-theoretical approach remains confusing in sociological discourse when "observation" is not provided with a sufficiently precise meaning.

## 2. Epistemological considerations

"Observation" can—for analytical reasons—be attributed to an "observing system" (Von Foerster, 1982). In an article entitled "Cybernetics of Cybernetics," Von Foerster (1979) proposed distinguishing first-order cybernetics as the cybernetics of observed systems (e.g., rockets) from second-order cybernetics as the cybernetics of observing systems (Baecker, 1996; Glanville, 2002). This formulation can be considered as the culmination of Von Foerster's work in the Biological Computer Laboratory at the University of Illinois, of which he had been the director for many years (Von Foerster, 1975; Varela & Goguen, 1978).

Shortly before this date, Maturana (1978) had proposed defining an observer in the context of developing autopoietic systems theory. Note that Maturana explained the generation of an observer *operationally* as a consequence of the biological process of autopoiesis in living systems and not—like Von Foerster—as an epistemological assumption underlying observation as an analytical category. Thus, this turn leaves Kant's (1781) epistemological concept of the observer as a transcendental subject more definitively behind because the observer is now grounded in the biological domain. Consequently, the focus is on the biological constraints and dynamics of an observer (Edelman, 1989). Maturana formulated his definition of an observer as follows:

> The magnitude of this recursive ontogenic structural coupling in any particular organism depends both on the degree of structural plasticity of its nervous system and on the degree to which the actual structure of its nervous system at any instant permits the occurrence of distinct relations of relative neuronal activity that operate as internal structural perturbations. When this takes place, even in the slightest manner, within the confines of a consensual domain, so that the relations of neuronal activity generated under consensual behavior become perturbations and components for further consensual behavior, an observer is operationally generated. (Maturana, 1978, at p. 49)

In other words: two variations can disturb each other as depicted in Figure 1. When the disturbance is structural and thus repeated over time, the interface can develop into a third system (e.g., a synapse). In other words, the uncertainty in the overlap can first be considered as a transmission between two variations. This co-variation can develop into a co-evolution over time. This third dimension was considered by Varela (1975, at p. 7) as the emergence of an autonomous state which can be distinguished from the two states that went into the process.

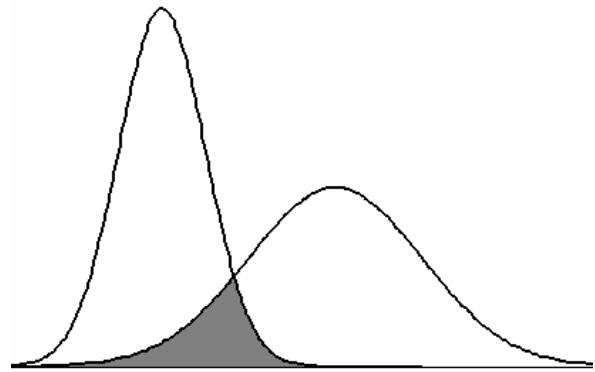

**Figure 1**
Two variations disturbing each other generate the possibility of selection of a signal by 'lock-in' when the disturbance is repeated over time.

The mechanism of how noise can become a signal in a selection environment can be specified by using Arthur's (1989, 1994) model for the "lock-in." A threshold can be passed in a random walk by chance and thereafter the process is self-reinforcing if the inscription of the signal leaves traces that can be further developed (Leydesdorff, 2001a). This process of lock-in and inscription occurs on both sides of the co-variation, a co-evolution can be developed as in a process of "mutual shaping" (McLuhan, 1964). This evolutionary perspective, however, is not addressed in Spencer Brown's (1969) *Laws of Form.* Varela (1975) elaborates on Brown's focus on two states: the 'marked' and the 'unmarked' ones. In a much later discussion, Brown (1994, at p. 51) added the possibility of an evolutionary process, but he formulated this mathematically as only an oscillation between the states:

> Similarly, when we get eventually to the creation of time, time is what there would be if there could be an oscillation between states.

The conditional is used by the author because one needs the additional assumption of, for example, evolution theory that an oscillation between two subsystems can be expected over time. The recursive operation (i.e., the clock of the system), however, is created on grounds external to the *Logic of Forms* (Spencer Brown, 1969), notably on the basis of substantive theorizing (Varela, 1975, at p. 20). The evolutionary theorist is not interested in the momentary appearance of an observer, but in its existence over time in terms of specific operations (Günther, 1967; Varela & Goguen, 1978).

Maturana (1978) noted that the behavior of an "observer" over time cannot be distinguished from the establishment of a semantic domain. He formulated this conclusion as follows:

> In still other words, if an organism is observed in its operation within a second-order consensual domain, it appears to the observer as if its nervous system interacted with internal representations of the circumstances of its interactions, and as if the changes of state of the organism were determined by the semantic value of these representations. Yet all that takes place in the operation of the nervous system is the structure-determined dynamics of changing relations of relative neuronal activity proper to a closed neuronal network. (*Ibid.*, at p. 49)

Thus, Maturana's observer remains completely endogenous to the system(s) under observation. The behaviour of this system can be described by an external super-observer in terms of an observer emerging within the system. Alternatively, one can describe this same effect as a semantic domain. The two descriptions are identical in the sense that they cannot be operationally distinguished at the level of the system itself. The emergence of a semantic domain is depicted in Figure 2. The overlap can generate a third domain within the system when the interaction is sustained by the two interacting systems over the time axis. The variations select upon each other mutually and thus are able to select certain variants for stabilization over time from the previous selections.

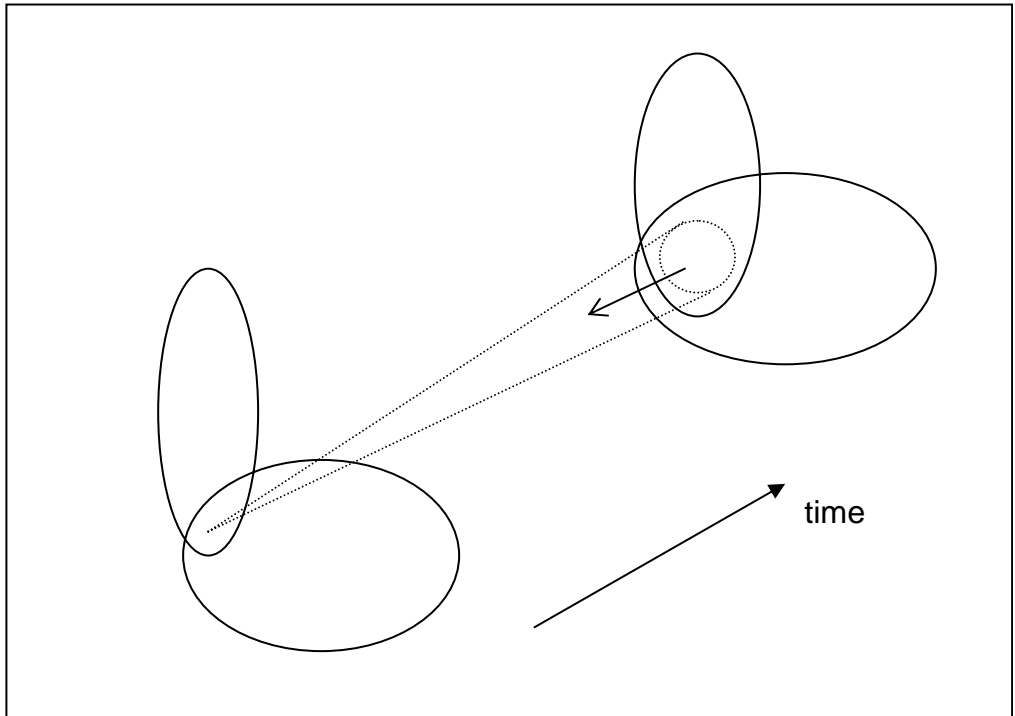

**Figure 2.** The structural coupling of the disturbance terms develops a semantic domain, that is, a first-order observer. Within the semantic domain a second-order observer cannot be distinguished *operationally* from a linguistic domain.

The mutually selected variants can be considered as providing a second-order variation. I have added to this picture in Figure 2 the next step that Maturana also took, namely the development of a language as the second-order operation of the semantic domain. This selection on the second-order variation can also be considered as a second-order selection or feedback. According to Maturana, "denotation" is the necessary requirement for the *linguistic* exchange. Denotation arises "only in a metadomain as an a posteriori commentary made by the observer about the consequences of operation of the interacting system" (*ibid.*, at p. 50).

In other words, language can be considered as an operation within the semantic domain that enables a first-order observer to enter into symbolic communication (e.g., by using words). The denoted entities can then be "discussed" regardless of whether these entities still exist in the domain from which they originated. While the coevolution adds time as a third degree of freedom to the two interacting variations, this next-order globalization requires one more degree of freedom in a system. A self-organizing system, therefore, can be modeled in a four-dimensional (hyper)space (Leydesdorff, 1994, 2001b).

One should keep in mind that "language" is here defined in terms of a system's operation and is not yet identified with "human language". Maturana's "biological language" is used, for example, among molecules that are structurally coupled in an autopoietic process of self-reproduction. Maturana carefully distinguished this "language" from the language used by biologists as supra-observers who entertain a biological discourse. In the biological case these two levels can clearly be distinguished. Sociology, however, has a more difficult task in this respect.

Maturana (1978, at p. 49) drew a radically constructivist conclusion from this specification of the biological discourse as a meta-language by stating that "one cannot speak about whatever one cannot talk." The human cognitive domain and its languaging are operationally closed (Maturana, 2000). However, it follows from this generalized definition of language as an operation that the domain of human language can only be generated *within* one semantic domain or another. The linguistic domain adds a reflexive layer of meaning processing to the semantic layer. The latter layer processes information as noise insofar as this information is not yet reflected and provided with by its second-order domain.

Let me anticipate on my conclusion that human language can be considered as specific in achieving one degree of freedom more than biological languages such as exchanges among molecules or animals. This degree of freedom enables us to develop a structural coupling between the (Shannon-type) information exchanges and the meaning exchanges. For example, some information can be provided with meaning, while the remainder is considered as noise. The same events can be provided with different meanings, and these meanings can again be communicated. The communication can again be noisy.

In the case of biological languages, the linguistic domain cannot further be developed as a communication system that feeds back as another operation on its own semantic domain. The "meanings" of the biological communications are provided in a linguistic domain, but these meanings changes among, for example, the cells according to the biology of the system. The life-cycles of the biological systems thus fix the meanings at each moment of time. These exchange processes are then changed by wear and tear. One would need a super-observer to reconstruct the biological domain as in molecular biology or biotechnology. However, the super-observer uses an angle of reflection different from that of the biological system under study. The neuronal system operates naturally within its own domain and generates a language that remains embedded (since naturalistically given).

In other words, a biological system has to adapt—at the risk of extinction—because the selection pressures are "natural." Selection environments, however, may already vary among ecological niches. Biological exchange relations can therefore experience additional selection pressure. However, the two types of selection pressures cannot be distinguished by the agents living in these niches unless these carrying agents would have the reflexive capacity to entertain concepts for mapping their situation. When these concepts can be communicated as denotations, a linguistic domain is generated operationally on top of the semantic domain. The system of communications can then become self-organizing and autopoietic in the sense of increasingly controlling and fine-tuning its own reproduction. This happens in biological cultures like insect populations.

Human language extends the biological concept of a linguistic domain because order is not constructed in a biological environment and then stabilized, but remains flexible and under construction as an expectation of order by using language and next-order (e.g., symbolically generalized) media of communication. The constructed order can be changed by a next-order system or at a next moment in time, i.e., by adding reflexively a new dimension to the system. Note that "reproduction" in this case no longer means reproduction in the biological sense, but rather the further development of the communication through the system's previously organized retention mechanisms. The lower-level systems upon which the next-level systems build can be innovatively reconstructed given the additional degree of freedom provided by linguistic exchanges. The represented and the representing systems may then begin to co-evolve. This co-evolution can be sustained by codes at the symbolic level.

## 3. The double hermeneutics in inter-human communication

In human interactions people are able to act as participant-observers and/or as external observers (Giddens, 1976). These two roles span other domains and roles for observers. However, the distinction between these two roles is analytical. In inter-human communication, this degree of freedom can be handled reflexively. The codification of this reflexivity spans the domain of sociological discourse (Giddens, 1979, 1984).

In other words, the "emic" perspective can be distinguished from the "etic" perspective by using a reflexive discourse (Geertz, 1973). One might even say that the sociological analysis emerges from reflexivity about this distinction. Under what conditions and precisely how can an analyst add knowledge to the knowledge of the participants? Can this knowledge be codified in a discourse which is different from and reflexive to the discourse of the subjects under study?

By focusing mainly on the meaning processing of the social system from an "etic" perspective, Luhmann (1984, 1997a) could make a number of important steps forward. For example:

- the possibility of functional differentiation of the communication. Since the communication system is considered as a system different from the carrying action systems (which Luhmann preferred to call "consciousness systems" in order to stress the priority of meaning processing; cf. Wasser, 2001), the differentiation can be expected to carry functions for the social system that differ from those functional for action. Thus, the four-function paradigm developed by Parsons (1937, 1951) for social *action* systems could be abandoned. The social system may use another alphabet of functions for the communication (Simon, 1969).
- These functions are carried by the symbolically generalized media of communication. These media operate by codifying the communication with potentially different meanings. Thus, using natural languages can be considered as a "primitive" (since natural) form of communication, notably, one that is not yet codified by using symbolic media. Prices, for example, enable us to speed up the economic process with an order of magnitude on compared with bargaining at a market.
- Consequently, communication and language can be further distinguished. Within scientific discourses, for example, communication is governed by code which determines whether or not the communication can be considered as "true." The code of communication closes the specificity of a scientific paradigm and thus a next-order jargon can be developed. This distinction between codified communication and language enables us to explain, for example, why and when a difference between "restricted discourse" and "elaborate discourse"—as distinguished in discourse analysis (Coser, 1975; Bernstein, 1971)—can be expected (Blauwhoff, 1995; Leydesdorff, 2002).
- In addition to the functional differentiation, Luhmann (1975, 2000) suggested a systemic differentiation in social communication between the levels of interaction, organization, and society. While the first two levels can be considered as specific formats for the integration, the level of society can evolve into a dynamics of functional differentiation and the self-organization of meaning processing. The dynamics between the functional and the institutional layer drive the system potentially in the higher gear of a knowledge-based system (Leydesdorff, 2001b, 2003).

Let us now return to the question of the status of the observer and his/her observation. Luhmann (1990a, 1993) followed Von Foerster in generalizing the concept of "observation" to the operation of a second-order system. As noted, the attribution of the observation to an observer was made by Von Foerster for the purpose of extending evolutionary theorizing in biology to meta-biological systems. However, Luhmann attempted to generalize this attribution also in terms of mathematics by claiming support from George Spencer Brown's (1969) *Logic of Forms*.

Spencer Brown formulated on the last page of this study as follows:

> An observer, since he distinguishes the spaces he occupies, is also a mark (…) In this conception a distinction drawn in any space is a mark distinguishing the space. Equally and conversely, any mark in a space draws a distinction.
>      We see now that the first distinction, the mark, and the observer are not only interchangeable, but, in the form, identical. (Spencer Brown, 1969: 76).

In an extensive note Spencer Brown (1969: 84) explained that the observer in this case does not have to be a human being. It can also be an animal. Thus, Spencer Brown was not yet specifying at the level of social systems theory, and not even at the level of biological evolution theory. The observer remained unspecified in terms of its operation because a "distinction" is drawn at a specific moment in time.

As noted, Varela (1975) extended Spencer Brown's logic to a calculus for self-referential systems by specifying the notion of identification of the distinction from a biological perspective. He speculated (at p. 22) that this solution at the biological level might provide a sound basis for a general theory of self-referential systems. Luhmann (1993), however, defined the observer *a priori* at the level of general systems theory. This has led to confusion between the definitions in terms of general systems theory and the specific distinctions and designations possible at the level of social systems.

Varela (1975; cf. Goguen & Varela, 1979; Varela & Goguen, 1978) elaborated on Spencer Brown's definitions in order to develop a calculus of self-reference. These authors argued that an "indication" adds to a "distinction" by marking one of the two distinguished states as primary; for example: "this," "I," "us," etc. The identification can be considered as the very purpose of the distinction. However, the (mathematical) distinction provides a necessary, but not a sufficient condition for an indication (Spencer Brown, 1969, at p. 1).[1] Indication requires the translation of the mathematical generality (of a distinction) into the contingent domain of substantive theorizing (for the identification). The identification completes the generation of an observer by using a second dimension. This accords with Maturana's (1978) definition of the (momentary!) generation of an observer. As noted by Varela (1975), the extension of the system with a third (temporal) dimension makes an autonomous system possible.

On the basis of this consensus Maturana & Varela (1980) could fully develop the biological theory of autopoiesis into an epistemology (cf. Luhmann, 1990a, at p. 81). Spencer Brown's observer is considered only a distinction, i.e., an instance ("mark") at one specific moment in time. It remains part of a logic or an arithmetic. An arithmetic was defined by Brown as "a calculus in which the constants operated on all have specific values" (Varela & Goguen, 1978, at p. 296). Maturana & Varela's observer, however, is part of an evolutionary process, i.e., it can be expressed as an algorithmic operation. The value of the variables can then change over time, while the variables remain identifiable (dx/dt). Because of the addition of selection over time, the variation is no longer subjected to a single selection mechanism operating at each specific moment in time. A relative stabilization of two selection mechanisms operating upon each other as mutual disturbances becomes possible. It has been argued above that some selections can then be selected for stabilization as an observer.

Note that Maturana & Varela's observers are defined as first-order observers because they are entrained in a life-cycle. The "observing systems" of Von Foerster, however, could be second-order observers which are not necessarily alife. (As noted above, Maturana distinguished additionally between a first-order semantic and a second-order linguistic domain.) Luhmann (1993) denied this distinction at the epistemological level because he wished to ground social systems theory as a substantive theory like biology in "observation" as an abstract operation defined at the level of *general* systems theory:

> The second-order observer, mind you, is a first-order observer as well, for he must distinguish and designate the observer he intends to observe. (Luhmann, 1993, at p. 20).

This deliberate blurring of the levels is fully consistent with Luhmann's proposal to consider the "observer" as a universal category that would provide a fundament to both second-order systems theory and the theory of social systems. For example, Luhmann (*ibid*.) could therefore formulate that "on the level of second-order observation it is then possible to see and state that …" However, communication systems do not have eyes for seeing and mouths for stating. Luhmann's texts tend to oscillate between first-order and second-order observers and this confusion generates what he indicates as a paradox.[2]

It is argued here that first-order and second-order observing systems operate differently, and therefore can be expected to entertain substantively different perspectives. "Observing systems" do not have to be identified as "observers" in a biological (or psychological) sense. The only requirement is that observing systems have to be able to carry "distinctions and designations." Systems other than identifiable "observers" may be able to do this. Social systems cannot be considered as biological observers because they remain distributed. Thus, they cannot be expected to perform bodily operations like seeing and speaking.

Luhmann himself sometimes used the word "dividuum" in order to emphasize the difference between a social system and *in*dividual observers. However, he defined "observation" as the general operation of drawing a distinction *and* making an identification. Thus, he extended the biological definitions of Maturana & Varela to a general systems notion that would be applicable to social systems as well, because he erroneously believed that the general definition had been provided by Spencer Brown in mathematical terms. However, this definition was a static one, whereas one needed Varela's biological extension for the evolutionary concept. The generalization of the notion of an observer—however abstract— into the sociological domain led eventually to a confrontation with Maturana. Maturana (1990) insisted on the physical existence of an observer. Von Foerster (1999) similarly expressed his reservations about Luhmann's generalizations.

*3.1     Can a second-order observer be considered a meta-observer?*

By putting the "observer" in social systems theory on a par with the abstract definition at the level of general systems theory a number of problems was generated. According to Luhmann's own theory of social systems, the social system would have to be an observer that is qualitatively different from (since orthogonal to) human observers. However, by defining this second-order observer in terms of the biological metaphor of a first-order observer, the second-order observer could also be considered as a supra-observer and thus a meta-biological metaphor was invoked. The second-order observer could be endowed with capacities (like "seeing") that tend to anthropomorphize this abstract notion. The Greek gods seemed to be back on stage. In the Luhmannian tradition, the consequent confusion is attributed to the limitations of language and discussed in terms of paradoxes (e.g., Fuchs, 2004). However, the paradoxes are not a result of linguistic limitations, like a tendency to reification in the German language, but the consequence of a category mistake at the level of the definitions which can be hidden by using linguistic metaphors.

This confusion of general (mathematical) and specific (biological) models has led to a stagnation in the research program of social systems theory during the 1990s. Gumbrecht (2003), for example, argued in a critical appraisal of Luhmann's legacy that the focus on the epistemology of paradoxes led to discussions during the 1990s that failed to add new insights to the well-known problem of the hermeneutic circle in the philosophy of language. This circle was not broken by distinguishing, for example, between a semantic (first-order) and a linguistic (second-order) domain in terms of different operations.

A distinction by a distributed system remains distributed and therefore cannot be identified other than as an uncertainty. An identification, however, would assume an identifiable observer like the analyst him-/herself. A distribution can only designate such distinctions in a *subsymbolical* mode. The uncertainty remains and cannot be resolved as in the case of an observation. The notion of "observation" as a concept of general systems theory—including trans-individual observations—and the attribution of the observation to a conscious observer would generate confusion at this level because the operation of the social system can lead to changes in the semantic fields among human beings beyond human will or intention. Such changes in the communication can be accessed by a reflexive observer, but they occur behind their backs as the non-linear results of their *and others'* interactions (Bhaskar, 1998; Marx, [1857], 1974: 176).

*3.2    Expectations, observational reports, and observations*

The specification of an uncertainty by making a distinction and an identification can be considered not yet as an observation in sociological discourse, but as the specification of an expectation. The distinction and designation specify a category, but not yet the value to be given to this category in the research process. One needs additionally a (scientific) observation for the specification of the latter. Thus, Luhmann used an epistemological notion of observations as empirical givens (from biology) that contradicts the concept of observation in empirical research in the social sciences.

For example, if one correlates two variables in a social science design, one is able to calculate whether the observations are statistically significant by using, for example, chi-square statistics. The values for the margin totals, that is, the nominal categories serve us for the specification of expected values, but empirical measurement provides the observed values. This difference between observed and expected is fundamental to the sociological design also in qualitative case studies. The cases that have occurred could have been different because of chance factors. In the qualitative case one confines the measurement to only the nominal level—that is, the description of the events in words—but the sociological design is thoroughly analytical and should avoid historicism (Weber, 1917; Popper, 1967). Case studies can only inform analytical theorizing because they provide the discourse with observational reports.

Sociological discourse processes observational reports rather than observations (Pask, 1975; De Zeeuw, 1993; cf. Matsuno, 2003). While the observations can be attributed to observers or more generally to agency,[3] only observational reports can function in the communication. In scientific communications, for example, the observational reports can be validated, i.e., provided with meaning by using the code of the scientific communication system. Are the observations reported also robust? Are they valid and reliable? A set of criteria can be applied and further developed within the scientific communication system used for the evaluation of knowledge claims in articles submitted for publication. The knowledge claims expressed in a linguistic domain provide the variation to a next-order codification in a symbolic domain.

In summary, the use of the word "observation" for the operation is not sufficiently reflexive in the case of the social system because this system operates in terms of expectations that can be informed by observational reports on the basis of observations by individuals. Observations update the expectations at the level of agency, but the communication requires a reflexive turn. I elaborated this above in the case of scientific communications, but similarly, in the economy, prices can be considered as expectations of the values of goods. Like other expectations entertained at the level of the social system, prices can be updated by observations if the reports about the observations can be appreciated at the level of the social system. The social system provides value to observations by codifying some observational reports more than others.

On the other side of the divide of the structural coupling between social systems and agencies value is provided to expectations by measurement because human beings can perceive a (natural or social) state of affairs by using sense data. Thus, the mechanisms for generating value are different in social systems and in consciousness systems. This distinction is obscured when "observation" is used as a general denominator for both expectation and observation, for the generation of value in terms of observations and in terms of coding (e.g., true/false), and for the generation of an observer.

If we distinguish these categories, it becomes clear that the meta-observer fails to exist at the level of the social system because this system remains distributed. The social system should not be anthropomorphized by using a biological metaphor (Hollak, 1963). This system has no organs for making an observation, but it is able to entertain substantive distinctions and designations in a distributed mode and over time. Luhmann (1984) himself emphasized these three dimensions of the communication: a communication is at the same time substantive, social, and temporal. The complex system which emerges from the nonlinear dynamics of three interacting subdynamics, is no longer necessarily grounded, that is, stabilized. In this case, the distinctions and designations function as expectations at the various levels of the system's organization as in interactions, organizations, and at the (potentially globalized) level of society (Luhmann, 1997b).

## 4. Conclusions

I have argued that a biological flaw in social systems theory has blocked its heuristic value in empirical research. The notion of identifiable observation was not sufficiently distinguished from that of a reflexive expectation. While the Luhmannian perspective has offered an elaboration of the etic-perspective in anthropology and sociology—because it abstracts from the intentions and the attributions of individual agency—it has remained tied to the biological origins of systems theory because of the generalization of the biological metaphor of an observer.

This generalization was justified in terms of Spencer Brown's (1969) *Logic of Forms*. However, one needs Varela's (1975) *biological* addition of "identification" as a necessary condition for "observation" for carrying the inference. Spencer Brown (1969) discussed "distinction" as a necessary condition for "identification," but only the former can be considered as a mathematical operation. Identification is substantive: it provides meaning to a distinction. Varela's evolutionary perspective, for example, could provide this substantive reflection. At the level of the social system, however, observations are not relevant for the system without further reflection in an observational report. These reports provide a distribution of observations that can be compared (and sometimes tested for their significance). The distribution is relevant for the social system because this system can operate only in terms of distributions.

Distributions at each moment in time can be expected to contain information. A redistribution of the distributed substance over time communicates an information or—in other words—generates a probabilistic entropy (Shannon, 1948). Thus, the perspective of social systems theory can be combined with the tools available in the mathematical theory of communications (Leydesdorff, 1995). However, this requires that what are considered as "observations" in biological systems theory be recognized reflexively and redefined as "expectations" for epistemological reasons.

Furthermore, this change in perspective from observations to expectations enables us to bridge the gap with the study of meaning processing in symbolic interactionism. Meaning is processed reflexively, that is, in terms of expectations (Knorr-Cetina & Cicourel, 1981). However, these expectations are structured and therefore systemic (Glaser, 1992; Lazarsfeld, 1995). In this tradition because of its focus on interactions, the systemic dimensions are indicated only as referentials. The deontologization of social systems theory intended by Luhmann (e.g., 1984, at p. 243) accords with this perspective (Gibson, 2000),[4] but the meta-biological metaphor has disturbed the translation of the heuristic value of social systems theory in sociological discourse (Giddens, 1984, at p. xxxvii; Leydesdorff, 1993).

## 5. Implications

Biological systems theoreticians are in a position different from that of social systems theoreticians because the biological systems are usually observable. Biological systems can be considered as "natural," and therefore the biologist is inclined to begin with the specification of an observable variation rather than the uncertainty of an expectation. As Maturana & Varela (1980: 90) formulated it emphatically:

> Notions such as coding and transmission of information do not enter in the realization of a concrete autopoietic system because they do not refer to actual processes in it.

While these authors insisted on the biological realization of "actual processes," Shannon's co-author Weaver (1949: 116f.) noted the problem of defining "meaning" from a mathematical perspective. The epistemological challenge of sociology is to abstract from observers as biological systems and to study systems of communication that are able to communicate about meanings and expectations in addition to observations (Leydesdorff, 2003).

Discourse does not require the realization of an observer. An identifiable observer would be only one among its possible realizations. The self-organization of meaning, however, can be considered as different from the self-organization of living (Luhmann, 1986). The systems-theoretical concept of "distinction and designation" can be used in sociological discourse only if the result of this operation is accorded the epistemological status of an "expectation." Given this reflection, the proof of the pudding remains the explanatory power of the specific perspectives of sociocybernetics and social systems theory among other sociological discourses (Leydesdorff, 1997). By specifying second-order observations as expectations, social systems theory and sociocybernetics can combine the constructivist with an empirical approach.

Weber, M. (1917, ³1968), "Der Sinn der 'Wertfreiheit' der Soziologischen und Ökonomischen Wissenschaften", in *Gesammelte Aufsätze zur Wissenschaftslehre,* Mohr, Tübingen, pp. 489-540.

---

**Notes**

[1] The reasoning is analogous to the information theoretical statement that a distribution contains an uncertainty. The specification of this uncertainty, for example, in terms of bits of information is yet content-free. The specification of a system of reference for the information can provide this information with meaning (Leydesdorff, 2003).
[2] See Kauffman's (2001) argument that mathematical reasoning about observers necessarily leads to paradoxes.
[3] Agency may also include the 'institutional agency' of a research group as an aggregate of agents or 'principal agency' when underlying agents are represented.
[4] 'The point from which all further investigations in systems theory must begin is therefore not identity but difference.

This leads to a radical de-ontologizing of objects as such—a discovery that corresponds to the analyses of complexity, meaning, the pressure to select, and double contingency.' (Luhmann, 1995, at p. 177).